# All-optical reservoir computing

François Duport,[1*] Bendix Schneider,[2] Anteo Smerieri,[1] Marc Haelterman,[1] and Serge Massar[3]

[1] *Service OPERA-photonique, Université Libre de Bruxelles (U.L.B.), 50 Avenue F. D. Roosevelt, CP 194/5, B-1050 Brussels, Belgium*
[2] *Department of Information Technology, Gent University, Sint-Pietersnieuwstraat 41, 9000 Gent, Belgium*
[3] *Laboratoire d'Information Quantique, CP 225, Université Libre de Bruxelles (U.L.B.), Boulevard du Triomphe, 1050 Brussels, Belgium*
[*] *Francois.DUPORT@ulb.ac.be*

**Abstract:** Reservoir Computing is a novel computing paradigm that uses a nonlinear recurrent dynamical system to carry out information processing. Recent electronic and optoelectronic Reservoir Computers based on an architecture with a single nonlinear node and a delay loop have shown performance on standardized tasks comparable to state-of-the-art digital implementations. Here we report an all-optical implementation of a Reservoir Computer, made of off-the-shelf components for optical telecommunications. It uses the saturation of a semiconductor optical amplifier as nonlinearity. The present work shows that, within the Reservoir Computing paradigm, all-optical computing with state-of-the-art performance is possible.

## References and links

1. H. Jaeger: *The "echo state" approach to analysing and training recurrent neural networks*, Technical Report GMD Report 148, German National Research Center for Information Technology (2001).
2. H. Jaeger: *Short term memory in echo state networks*, GMD Report 152, German National Research Institute for Computer Science (2001).
3. W. Maass, T. Natschläger, and H. Markram: "Real-time computing without stable states: a new framework for neural computation based on perturbations," Neural Comput. **14** (11), 2531-2560 (2002).
4. H. Jaeger and H. Haas: "Harnessing nonlinearity: predicting chaotic systems and saving energy in wireless communication," Science **304** (5667), 78-80 (2004).
5. J. J. Steil, "Backpropagation-decorrelation: online recurrent learning with O(N) complexity," in *Proceedings of IEEE International Joint Conference on Neural Networks* (IEEE, 2004), 843–848
6. R. Legenstein and W. Maass: *New directions in statistical signal processing: from systems to brain*, (MIT Press 2005), Chap. "What makes a dynamical system computationally powerful?" 127–154.
7. D. Verstraeten, B. Schrauwen, M D'Haene, and D. Stroobandt: "An experimental unification of reservoir computing methods," Neural Networks **20**(3), 391-403, (2007).
8. W. Maass, P. Joshi, and E. Sontag: "Computational aspects of feedback in neural circuits," PLOS Comput. Biol. **3**, 1–20, (2007).
9. H. Jaeger, M. Lukosevicius, D. Popovici, and U. Siewert: "Optimization and applications of echo state networks with leaky-integrator neurons", Neural Networks **20**, 335–52, (2007).
10. D. V. Buonomano and W. Maass: "State-dependent computations: spatiotemporal processing in cortical network," Nat. Rev. Neurosci. **10**, 113–125, (2009).
11. M. Lukoševičius and H. Jaeger: "Reservoir computing approaches to recurrent neural network training", Computer Science Review **3**, 127-149 (2009).
12. B. Hammer, B. Schrauwen, and J. J. Steil: "Recent advances in efficient learning of recurrent networks", in *Proceedings of the European Symposium on Artificial Neural Networks*, 213–216, (2009).
13. F. Triefenbach, A. Jalalvand, B. Schrauwen, and J. Martens: "Phoneme recognition with large hierarchical reservoirs," in *Proceedings of Advances in Neural Information Processing Systems* **23**, 1–9, (2010).
14. F. Wyffels and B. Schrauwen: "A comparative study of reservoir computing strategies for monthly time series prediction", Neurocomputing **73** (10-12), 1958-1964, (2010).
15. M. Lukoševičius, H. Jaeger, and B. Schrauwen: "Reservoir computing trends", KI - Künstliche Intelligenz, pp. 1-7, 2012
16. C. Fernando and S. Sojakka: "Pattern recognition in a bucket," in *Proceedings of the 7th European Conference on Artificial Life* **2801,** W. Banzhaf, J. Ziegler, T. Christaller, P. Dittrich, and J. Kim editors, 588-597 (2003).

17. F. Schürmann, K. Meier, and J. Schemmel: "Edge of chaos computation in mixed-mode vlsi - a hard liquid," in *Proceedings of Advances in Neural Information Processing Systems*, L. K. Saul, Y. Weiss, and Léon editors (MIT Press, 2005).
18. Y. Paquot, B. Schrauwen, J. Dambre, M. Haelterman, and S. Massar: "Reservoir computing: a photonic neural network for information processing", Proc. SPIE **7728**, 77280B-1 (2010).
19. A. Rodan and P. Tino: "Simple deterministically constructed recurrent neural networks," *Intelligent Data Engineering and Automated Learning* (IDEAL, 2010), 267-274.
20. A. Rodan and P. Tino: "Minimum complexity echo state network", IEEE T. Neural Netw. **22**, 131–144, (2011).
21. L. Appeltant, M.C. Soriano, G. Van der Sande, J. Danckaert, S. Massar, J. Dambre, B. Schrauwen, C.R. Mirasso, and I. Fischer: "Information processing using a single dynamical node as complex system", Nat. Commun. **2**, Article number: 468 (2011).
http://www.nature.com/ncomms/journal/v2/n9/full/ncomms1476.html
22. Y. Paquot, F. Duport, A. Smerieri, J. Dambre, B. Schrauwen, M. Haelterman, and S. Massar, "Optoelectronic reservoir computing," Sci. Rep. **2**, Article number: 287, (2012).
http://www.nature.com/srep/2012/120227/srep00287/full/srep00287.html
23. L. Larger, M.C. Soriano, D. Brunner, L. Appeltant, J. M. Gutierrez, L. Pesquera, C.R. Mirasso, and I. Fischer: "Photonic information processing beyond turing: an optoelectronic implementation of reservoir computing", Opt. Express **20**, 3241-3249 (2012).
http://www.opticsinfobase.org/oe/abstract.cfm?URI=oe-20-3-3241
24. K. Vandoorne et al.: "Toward optical signal processing using photonic reservoir computing," Opt. Express **16** (15): 11182-11192 (2008).
25. K. Vandoorne, J. Dambre, D. Verstraeten, B. Schrauwen, and P. Bienstman: "Parallel reservoir computing using optical amplifiers", IEEE T. Neural Netw. **22**(9), 1469-1481, (2011).
http://www.opticsinfobase.org/oe/abstract.cfm?URI=oe-16-15-11182
26. J. Dambre, D. Verstraeten, B. Schrauwen, and S. Massar: "Information processing capacity of dynamical systems", Sci. Rep. **2**, Article number: 514, (2012).
27. H. Jaeger: "Adaptive nonlinear system identification with echo state networks," *Advances in Neural Information Processing Systems* **8**, 593-600 (2002).
28. V. J. Mathews, "Adaptive algorithms for bilinear filtering", Proc. SPIE **2296** (1), 317–327, (1994).
29. http://soma.ece.mcmaster.ca/ipix/dartmouth/datasets.html
30. D. Verstraeten, B. Schrauwen, and D. Stroobandt. "Isolated word recognition using a liquid state machine". In *Proceedings of the 13th European Symposium on Artifcial Neural Networks* (ESANN), 435-440, (2005).
31. Texas Instruments-Developed 46-Word Speaker-Dependent Isolated Word Corpus (TI46), September 1991, NIST Speech Disc 7-1.1 (1 disc), (1991).
32. R. Lyon: "A computational model of filtering, detection, and compression in the cochlea". in *Proceedings of IEEE International Conference on Acoustics, Speech, and Signal Processing*, 1282-1285 (1982).

# 1. Introduction

Reservoir Computing is a novel neural network computing paradigm introduced at the onset of the $21^{st}$ century [1-15]. It is very well suited to process time dependent inputs, and provides state-of-the-art performance on tasks such as speech recognition, time series prediction, etc. (see [11,15] for reviews).

A Reservoir Computer consists of a high-dimensional nonlinear dynamical system, the reservoir, driven by a time dependent input, and an output layer in which the output signal is calculated. The time dependent output of the reservoir is given by a linear combination of the instantaneous internal states of the reservoir. The weights of this linear combination are the only parameters of the system which are trained to match the reservoir output to a target sequence. In addition, a few parameters of the dynamical system are usually tunable, to adjust how close the system is to the edge of stability and to ensure adequate coupling between the internal variables. The simplicity and flexibility of the concept of Reservoir Computing makes it eminently suitable for experimental implementation. Indeed, given the loose constraints which a system must satisfy in order to constitute a suitable reservoir, a variety of experimental realizations have so far been built. These include using a bucket of water [16], an analogue VLSI chip [17], and a delay system with a single nonlinear node. The latter architecture was introduced in an experimental architecture in Ref. [18] and studied independently from a theoretical point of view in Ref. [19,20]. Because of its simplicity, this architecture has proven very powerful for experiments. It enabled, using an electronic implementation, the first experimental Reservoir Computer with performance comparable to state-of-the-art digital implementations [21], as well as the first optoelectronic implementation of Reservoir Computing [18,22,23]. For a further general discussion, we refer to the supplementary material of Ref. [22] in which a roadmap for building experimental reservoirs is presented.

The realization of all-optical Reservoir Computers is attractive. All-optical computation has proven a tantalizing but frustrating goal since decades. On the one hand, optics has great promise for computation, such as inherent speed and parallelism. On the other hand, the difficulty of realizing optical nonlinearities and of building complex optical architectures has hindered these efforts. Reservoir Computing seems a promising new avenue. Indeed, in Reservoir Computing one can directly use the available nonlinearities for computation, rather than first adapting them so as to implement specific operations (e.g. logical gates such AND, OR, etc…) which are then combined into more complex architectures. This implies that Reservoir Computing is highly flexible, a desirable feature when, as in optics, implementing nonlinearities is difficult.

The possibility of an all-optical implementation of Reservoir Computing based on an array of Semiconductor Optical Amplifiers (SOA) on a chip has been proposed through numerical simulations in Ref. [24,25]. Here we report the first all-optical experimental implementation of a Reservoir Computer. The architecture we use is based on a fiber optics delay loop with a single nonlinear node and off-line training [21-23]. The nonlinearity is provided by the saturation gain effect in a SOA.

The experiment reported here constitutes a significant step towards the possible development of analogue ultrafast all-optical computers. However, independently of the ultimate realization of this goal, the present work is also of fundamental interest as it illustrates the versatility and robustness of the delay loop architecture. Indeed, at the conceptual level our experiment differs from previous ones both in the type of nonlinearity used, and in the presence of significant amounts of noise that arise due to the spontaneous emission from the SOA.

In the next section of this paper we outline the operating principle of the Reservoir Computer and detail the choice we made for its all-optical implementation. We then describe in Sec. 3 how to operate the reservoir to achieve the best performances. And finally, in Sec. 4 we present the results obtained on a variety of benchmark tasks in order to evaluate these performances.

## 2. The photonic hardware implementation

### 2.1. Principles of reservoir computing

On a conceptual level, a Reservoir Computer can be thought of as a collection of internal nodes, whose states $x_i$ evolve in discrete time $n$ according to a nonlinear recurrent map of the form

$$x_i(n) = F_{NL}\left(\sum_{j=1}^{N}\alpha A_{ij} x_i(n-1) + \beta m_i u(n)\right), \quad i = 1, 2, ..., N \qquad (1)$$

where $N$ is the number of nodes, $F_{NL}$ is a nonlinear function, $u(n)$ is the input signal, $m$ is the input mask vector, and $A$ is the interconnection matrix. The feedback gain $\alpha$ and the input gain $\beta$ are parameters that are used to tune the dynamics of the reservoir in order to find its best working point. The input mask $m_i$ enriches the dynamics of the reservoir by distributing the same input to different nodes with different weights $m_i$.

The aim of a Reservoir Computer is to carry out computation on the input signal $u(n)$. The result of the computation is an output $\hat{y}(n)$. It is given by a linear combination of the node states

$$\hat{y}(n) = \sum_i W_i x_i(n). \qquad (2)$$

The output weights $W_i$ are calculated to minimize the distance between the actual output $\hat{y}(n)$ and the desired output $y(n)$. This is typically done by considering a "training" sequence for which the input and desired output are known. In many cases the distance between the actual and desired output is measured using the Normalized Mean Square Error (NMSE), given by

$$NMSE = \frac{\left\langle (y-\hat{y})^2 \right\rangle_n}{\left\langle (y-\langle y\rangle)^2 \right\rangle_n}. \qquad (3)$$

Note that the input and output can both have more than one dimension, in which case the input mask $m_i$ and readout weights $W_i$ become respectively the matrices $m_{ik}$ and $W_{ik}$, as will be considered in Sec. 4.5 and 4.6.

While in most digital implementations the connection matrix $A$ is randomly generated, there are in fact very few requirements on its structure, the main one being that its spectral radius multiplied by the feedback gain $\alpha$ should be less than 1 in order to avoid instabilities. It has been shown [19,20] that even very simple interconnection matrices $A$ can lead to excellent results. We can therefore choose an architecture where each node state $x_i$ only depends on the adjacent node state $x_{i-1}$. Such an interconnection matrix can be naturally implemented in the time-multiplexed sequential reservoir configuration described in the next section.

*2.2. Configuration with a delay loop and a single nonlinear node.*

The Reservoir Computer implementation using a single nonlinear node and delayed feedback has been described in detail in Refs [21-23]. Its main advantage is experimental simplicity, since it requires very few components.

To implement this architecture, we need to go from the discrete time $n$ used in eq. (1) to continuous time. To this end each input $u(n)$ undergoes a sample and hold procedure, during which it is held for a time $T$. Each interval of length $T$ is subdivided into $N$ intervals of length $\theta$. The i-th interval of duration $\theta$ $(i=1,...,N)$ is associated with the input value $m_i u(n)$, where $m_i$ is the corresponding value of the input mask. It is these values $m_i u(n)$ which are sequentially used to drive the reservoir.

We denote by $T'$ the round-trip time of the delay loop. Two approaches have been proposed to implement the single nonlinear node and delayed feedback scheme. In the first, used in Refs. [21,23], one takes $T = T'$. In this case, to enrich the dynamics, it is necessary to include in the delay loop a lowpass filter with time constant typically of order a few times $\theta$. This internal time-scale couples successive node states to each other.

In the second approach, used in Ref. [22], the nonlinearity is instantaneous, but one desynchronizes the input from the round-trip time, for instance by taking $T' = (N+1)\theta$. This introduces a coupling from each internal variable $x_i$ to the adjacent one $x_{i-1}$. It is this second approach that is used in the present experiment.

*2.3. Experimental implementation.*

Our experimental setup is depicted in Fig. 1. The all-optical reservoir is implemented using off-the-shelf fiber components operating in the telecommunication C-band. It uses an incoherent light source so that interference effects can be neglected and the internal variables $x_i(n)$ correspond to the light intensity inside the loop. The nonlinear feedback loop consists of a fiber spool (1.6 km of single mode fiber), an SOA (Covega C-Band SOA1117) and a tunable optical attenuator (Agilent 81571A). The latter determines the feedback strength $\alpha$ that appears in the eq. (1). In addition the delay loop contains an isolator to avoid counter-propagating Amplified Spontaneous Emission (ASE) within the SOA and a band pass filter (commercial Coarse Wavelength Divison Multiplexer centered on 1550nm) that removes a large part of the spontaneous emission noise from the SOA. The noise figure of this amplifier has been measured on the photodiode to take values between 25 dB and 45 dB, depending on the input power and the pump current.

The round-trip time of the loop is $T' = 7.9437\,\mu s$. With $N = 50$ internal variables and $T' = (N+1)\theta$, the input time-scale is $T = 7.7880\,\mu s$, and $\theta = 155.76\,ns$.

Nonlinearity in a Reservoir Computer is essential to fulfill complex tasks. In our system the nonlinearity is provided by the saturation of the optical gain in the SOA. In Fig. 2 we show the measured power amplification characteristics of the SOA for various injection currents. These characteristics constitute the nonlinear function $F_{NL}$ of our reservoir. As can be seen in Fig. 2, the injection current controls the shape of this nonlinear function and therefore constitutes an adjustable parameter in our system. For a given task, the injection current is varied to get the best performance. For most tasks we have found that operating in a relatively linear regime, corresponding to a low input current, provides best performance. Note that in our experimental conditions the amplifier is always used far from the complete saturation regime.

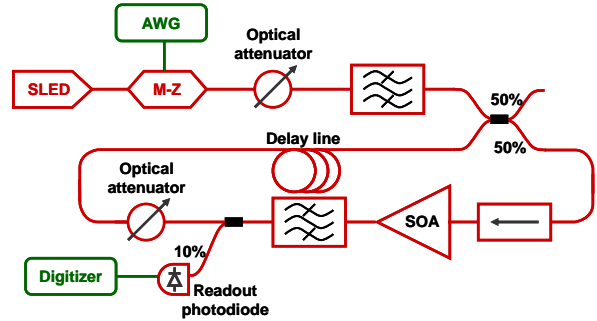

Fig. 1: Schematic of the experimental set-up of the all-optical reservoir. The reservoir consists of off-the-shelf fiber optics components operating in the telecommunication C-band. It is based on nonlinear all-optical loop operating in incoherent regime. Optical components are in red whereas electronic components are in green. The all-optical loop is driven by the input optical signal. A Superluminescent Light Emitting Diode (SLED) generates broadband white light. An electronic signal corresponding to the time dependent input multiplied by the input mask is generated by the Arbitrary Waveform Generator (AWG). This electronic signal drives an integrated Lithium Niobate Mach-Zehnder intensity modulator (MZ) thereby producing a time dependent input optical signal whose intensity is adjusted with a variable attenuator. The input optical signal is injected into the cavity by means of a 50/50 fiber coupler. The cavity itself consists of an isolator, a Semiconductor Optical Amplifier (SOA), a variable optical attenuator, and a fiber spool that acts as delay line. The cavity operates below the lasing threshold. A 90/10 fiber coupler is used to send 10% of the cavity intensity to a readout photodiode and then to a digitizer. Coarse Wavelength Division Multiplexers (CWDM, boxes with waves) are used to select a wavelength band near the maximum of emission of the SLED and the maximum gain of the SOA.

To drive the reservoir we use an incoherent light source (40 nm broadband Superluminescent Light Emitting Diode (SLED, Denselight DL-CS5254A) with peak emission at 1560 nm). The time-varying inputs are created by means of an Arbitrary Waveform Generator (AWG, National Instrument model PXI-5422) that drives a Mach-Zehnder Modulator (MZM, Photline model MXAN-LN-10). The intensity injected into the reservoir is chosen by means of a variable optical attenuator (Agilent, model 81571A). Two practically identical Coarse Wavelength Division Multiplexers (CWDM) with pass band [1540-1558.5nm] are added at the output of the SLED and the SOA. These filters improve the signal-to-noise ratio (SNR) by matching the bandwidth of the source to that of the cavity and removing a large portion of the ASE noise generated by the SOA.

A fraction (10%) of the circulating light intensity is extracted from the loop via a fiber coupler in order to record the evolution of the node states by means of a photodiode (TTI TIA-525 120 MHz bandwidth), and a digitizer (NIPXI-5124 Digitizer), whose output is stored in a desktop computer used for post-processing. The input and output are sampled by the AWG and digitizer at 200MSamples per second.

In summary, the present experiment is largely inspired by our earlier work Ref. [22]. In particular the AWG and digitizer, as well as most of the numerical code used to control the system, are identical. For this reason it is instructive, as we do in the following, to compare the two implementations. In the present all-optical implementation, the two variable optical attenuators allow independent adjustment of the feedback gain $\alpha$ and the input gain $\beta$. An important difference with respect to our previous work Ref. [22] is the shape of the nonlinearity (saturable gain versus sine nonlinearity). In the present case, the nonlinearity can be adjusted to some extent by modifying the injection current of the SOA, see Fig. 2. Another important difference is the spontaneous emission noise of the SOA that affects the SNR. An example of response of the all-optical reservoir to a simple input is given in Fig. 3, illustrating the fading memory of the reservoir and its nonlinear response.

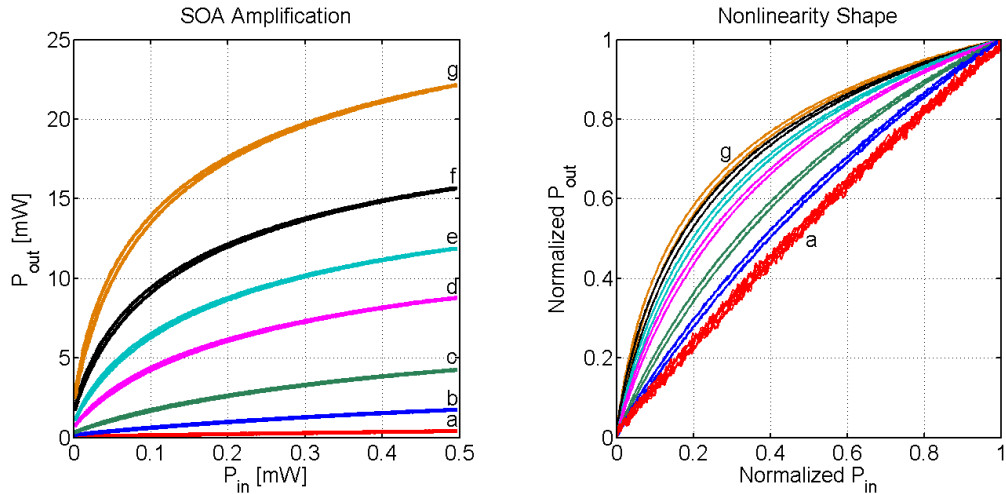

Fig. 2: Incoherent light amplification characteristics of the SOA. Left hand: output power versus input power. The SOA gain is almost linear at low input powers and bends over at input powers above the saturation level. The overall output power increases with increasing injection current. Due to spontaneous emission, the output power is non zero even at zero input power. From bottom to top: injection current is a) 160 mA, b) 200 mA, c) 240 mA, d) 300 mA, e) 340 mA, f) 400 mA, and g) 520 mA. (The hysteresis effect is due to the high-pass filter of the modulator's driver used for the record). Right hand: normalized input-output relation. The data is the same as in the left panel, but the input and output powers are normalized to lie in the interval [0,1]. It is this function that characterizes the nonlinearity of the reservoir, which can be tuned by changing the pump current.

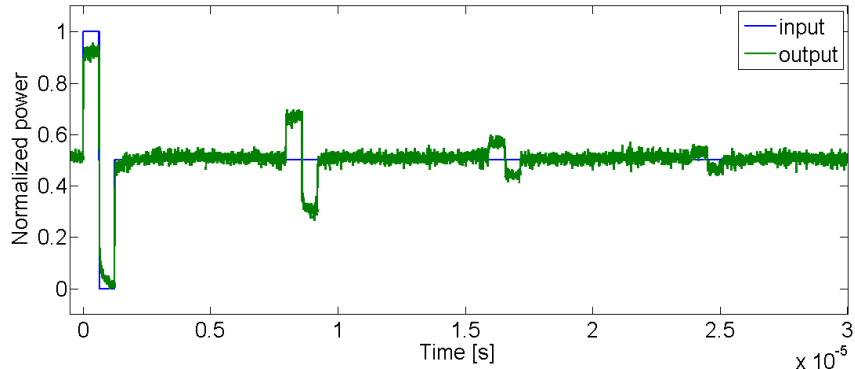

Fig. 3: Example response of the all-optical reservoir. The amount of light injected into the reservoir is kept constant (normalized value 0.5). At time t=0, the input signal (blue curve) is momentarily set to 1, then to zero, before returning to its initial value of 0.5. The response of the reservoir is recorded in green. The nonlinearity of the reservoir due to saturation of the SOA is clear from the initial response (positive response is less than negative response). The fading memory of the all-optical system is also clear. Experimental parameters are: injection current 187mA, feedback gain -1.6 dB, and average input power rescaled to its value at the entrance of the SOA 350.8μW (-4.55dBm).

## 3. Operation mode of the all-optical reservoir

### 3.1. Signal pre-processing

Because our implementation is based on light intensity coded information, signals need to be adapted so that they can be represented by a positive scalar regardless of the task under study. To do so, the masked input sequence, i.e. the product $m_i u(n)$ of the input mask and the input signal, is renormalized to lie within the interval [0,1]. It is then converted into a voltage signal that drives the Mach-Zehnder Modulator (MZM) in such a way that the intensity of the light after the modulator is $m_i u(n) P$ where $P$ is the maximum power available. This is achieved by applying to the MZM the arcsine of the normalized product $m_i u(n)$ and multiplying the result by a factor chosen such that the voltage signal spans the interval [0, $V_\pi$]. This procedure exploits the maximum dynamical range of the MZM while removing the nonlinearity that is potentially present due to the nonlinear transfer function of the MZM. This allows us to compare the performance of the all-optical reservoir with the previous experiments Refs. [21-23]. Note that if we do not carry out this pre-compensation, the performance of the reservoir can improve on certain tasks due to the added nonlinearity provided by the transfer function of the MZM.

### 3.2. Signal post-processing

In our experiment the reservoir training and testing are realized on the basis of the values recorded by the digitizer and are thus not performed in real time. The entire recorded intensity time trace is first rescaled to lie in the interval [-1,+1]. Then each internal state is measured by averaging the normalized intensity around the middle of the corresponding time window over a duration of $\theta/2$. By doing so, we discard the leading and trailing edges of each time interval where abrupt jumps in the light intensity take place. This amounts to eliminating most of the influence of the photodetector response time on the state values.

For each experiment the desired or "target" output sequence $y(n)$ corresponding to a given input sequence $u(n)$ is defined according to the task under consideration. The reservoir training then consists in applying a least mean square algorithm that calculates the set of weight values $W_i$ that minimizes the mismatch between the desired output sequence $y(n)$ and the actual output sequence $\hat{y}(n)$ as calculated from Eq.(2). In order to make the reservoir more robust against overfitting ridge regularization is used.

Once the training is performed, the weight values $W_i$ are kept constant so that the reservoir can be tested with the remaining of the recorded normalized intensity time trace. Note that besides the training and testing sequences, the recorded output intensity time trace also contains a "warm-up" sequence that is not used in the training phase. Its role is to ensure that the reservoir has no memory of the steady state from which it systematically starts in the various experiments. Its length is typically 2 to 4 times the number of nodes in the reservoir (typically 200-400 samples). The total length of the warm up, training, and testing sequences is usually around ten thousand samples.

### 4. Results

*4.1. Operating point*

We investigated several benchmark tasks that have been used previously in the Reservoir Computing community. In the following paragraphs we describe the different tasks and give the corresponding scores obtained with our all-optical Reservoir Computer. We also quote the operating point used to get best performance on the task. The operating point is specified by three parameters.

The injection current of the SOA determines the shape of the nonlinearity, see Fig. 2. It also determines the noise figure of the SOA due to its ASE. For most tasks we have found that a small injection current, corresponding to a rather linear regime, gives best performance. The value of the injection current used is quoted for all tasks. In most cases it is 187mA.

The feedback gain $\alpha$ is estimated by measuring the small signal gain of the SOA and taking into account the other losses in the cavity. The largest feedback gain, corresponding to minimum loss in the cavity, is -1.6dB.

The input gain $\beta$ is quantified by taking the power $m_i u(n) P$ at the output of the MZM, averaging over the mask values $m_i$ and the input values $u(n)$, and then propagating the resulting average power (taking into account losses in the variable attenuator and coupler) to the input of the SOA. The input power, brought back to the input of the SOA, is measured in dBm and µW, which allows comparison with the transfer function of the SOA, as reported in Fig. 2.

*4.2. Numerical simulations*

We have developed a simple numerical model of the experiment, based on integrating numerically the recurrence

$$x_i(n) = \begin{cases} F_{NL}\left(\alpha \cdot x_{i-1}(n-1) + \beta \cdot m_i \cdot u(n)\right) + v_i(n) & 2 \leq i \leq N \\ F_{NL}\left(\alpha \cdot x_{N+i-1}(n-2) + \beta \cdot m_i \cdot u(n)\right) + v_i(n) & i = 1 \end{cases} \quad (4)$$

where $F_{NL}$ is the measured response of the SOA given in Fig. 2, and $v_i(n)$ is the noise added to node variable $x_i$ at time $n$ (the $v_i(n)$ are independent indentically distributed Gaussian variables with zero mean). This numerical model takes into account the measured transfer functions of all the components. It does not take into account any bandpass effects of the electronics (either at the input or readout). The ASE noise added by the SOA is modeled by adding the noise terms $v_i(n)$. The aim of this numerical model was to establish whether the architecture used (single nonlinear node and delay loop) combined with the form of the nonlinearity (saturable nonlinearity) introduced fundamental limitations to our experiment. Except if relevant, we do not quote the predictions of these numerical simulations.

*4.3.Memory capacities*

The memory capacities characterize in a simple way how a reservoir processes information. The inputs $u(n)$ are taken to be independent identically distributed random variables. In the present work the $u(n)$ are drawn from the uniform distribution over the interval [-1,1].

The aim is to reconstruct a function $y[u]$ of the previous inputs. The quality of the reconstruction is measured by the capacity $C[y] = 1 - NMSE[y]$, where *NMSE* is the normalized mean square error, so that a perfect reconstruction of the input gives a capacity of 1, while a completely uncorrelated reconstruction gives a capacity of 0. Generally we consider a set of functions $y_k$, and the total capacity for this set: $C = \sum_k C[y_k]$.

In the case of the *linear memory capacity*, introduced by Jaeger in 2001 [2], the task is to reconstruct the input $k$ steps in the past. That is $y_k(n) = u(n-k)$. Upon summing the capacities over all delays $k$, one obtains the linear memory capacity.

The *nonlinear capacities* introduced in [26] are extensions of the linear capacity to the case where the function to reconstruct is a nonlinear function of the past inputs. Here we consider only second order polynomials of the past input. Namely we consider the second order Legendre polynomial of the input $k$ steps in the past $y_k(n) = 3u^2(n-k) - 1$ and the product of two inputs $k$ and $k'$ steps in the past $y_{kk'}(n) = u(n-k)u(n-k')$. Upon summing the capacities over $k$ (in the case of the second order Legendre polynomials) and over $k$, $k'$ (in the case of products) one obtains the quadratic memory capacity and the cross memory capacity, respectively.

A key result proven in [26] is that the sum of the linear memory capacity, the quadratic memory capacity, and the cross memory capacity is less or equal to the number of internal variables $N$ (in our case $N = 50$). This generalizes the result of [2] where it is shown that this inequality holds in the case of the linear memory capacity. Note that in order to saturate the bound one should in general add all higher order polynomials, see [26] for the conditions under which this bound is saturated. In other words, while the linear memory capacity quantifies how well the reservoir remembers past inputs, the quadratic and cross memory capacities quantify how well the reservoir can compute second order polynomials of the past inputs.

Table 1 presents the maximum value obtained for each of the memory capacities of our all-optical reservoir for a pump current of 187mA, and compares them to the same quantities for our previous optoelectronic experiment [22], while keeping the number $N = 50$ of internal variable constant. The total memory capacity mentioned in the last line is the sum of the linear, quadratic, and cross memory capacities. To compute all the values reported we considered all the contributions up to k=100, even if no significant contribution is expected to be found for $k > N$.

**Table 1: Linear, quadratic, cross and total memory capacities of the system with 50 internal variables and a pump current of 187 mA compared to memory capacities of an optoelectronic reservoir with the same number of internal variables. Each quantity is reported for the optimal choice of feedback gain $\alpha$ and the input gain $\beta$. The last line gives the sum of the first three quantities, once again maximized over $\alpha$ and $\beta$.**

|  | All-optical Reservoir Computer | Optoelectronic Reservoir Computer ([22]) |
|---|---|---|
| Max linear memory capacity | 20.8 | 31.9 |
| Max quadratic memory capacity | 4.16 | 4 |
| Max cross memory capacity | 8.13 | 27.3 |
| Max total memory capacity | 28.84 | 48.6 |

Fig. 4 shows how these quantities change in terms of the feedback gain $\alpha$ and the input gain $\beta$.

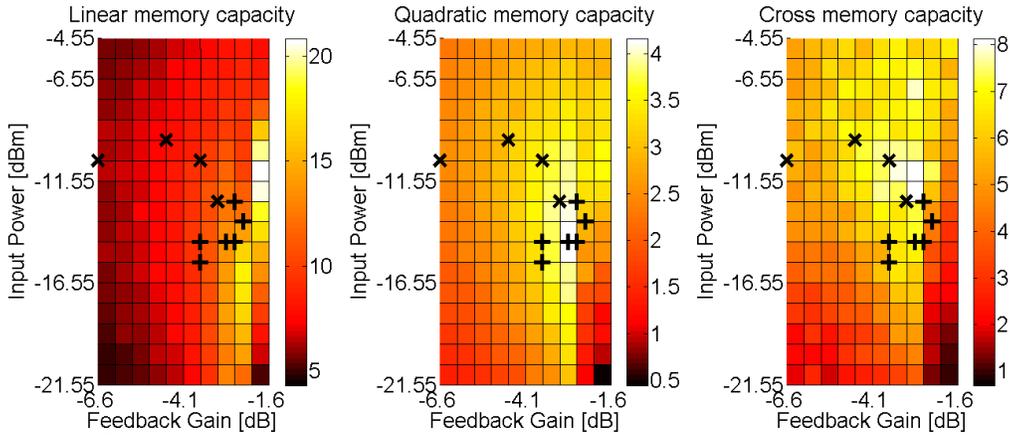

Fig. 4: Linear, quadratic, and cross memory capacities for the all-optical reservoir, as a function of the feedback gain feedback gain $\alpha$ and the input gain $\beta$ for an injection current of 187mA.

Note that the color scale is different for each capacity. The + signs locate the optimum working points for the channel equalization task at the different signal-to-noise ratios. The x signs locate the optimum working points for the radar task at low sea state over the 10 delays of prediction (some delays have identical optimal working points). This shows that the capacities are correlated to, but do not explain completely, the optimal working points.

Compared with the capacities of the optoelectronic architecture, the memory capacities of the all-optical reservoir are lower. In particular, we were unable to reach a large cross memory capacity or a total memory capacity close to its optimal value, i.e., the number of nodes in the reservoir. This is attributed to the ASE noise of the SOA that is inherent to our all-optical configuration and decreases the SNR.

*4.4.Channel equalization task*

We now apply the all-optical reservoir to a task related to telecommunication applications, first used in the context of Reservoir Computing by Jaeger [4]. It consists of channel equalization in a wireless communication link undergoing multi-path symbol interference as well as nonlinear distortion. The transmitted symbols $d$ are drawn randomly from the set $\{-3,\ -1,\ 1,\ 3\}$ and undergo the following channel transformation.

$$q(n) = 0.08d(n+2) - 0.12d(n+1) + d(n) + 0.18d(n-1) - 0.1d(n-2) + 0.091d(n-3)$$
$$- 0.05d(n-4) + 0.04d(n-5) + 0.03d(n-6) + 0.01d(n-7) \qquad (5)$$
$$u(n) = q(n) + 0.036q^2(n) - 0.011q^3(n) + noise$$

White noise is added to obtain a Signal-to-Noise Ratio (SNR) ranging from 12 dB to 32 dB. The detected signal $u(n)$ obtained at the end of the communication channel constitutes the reservoir input, while the target output is the original sequence $d(n)$.

This task is evaluated through the calculation of the Symbol Error Rate (SER), representing the percentage of misclassified symbols. The results we obtained for this task are presented in Fig. 5, together with the results obtained for the same task with our optoelectronic reservoir [22], and the ones obtained by running the numerical simulations described in section 4.2 of this paper. This figure reveals that the all-optical reservoir performs as well for a signal-to-noise ratio ranging from 12 to 16 dB, while for larger SNR the performance is slightly degraded with respect to the optoelectronic implementation. Note, however, that even in this high SNR range the measured SER is still one order of magnitude better than the bilinear filtering technique presented in [28]. It should also be noted that there is a good agreement between the results of the optical experiment and the ones obtained with the numerical simulations, using a distribution of the internal noise $v_i(n)$ with a standard deviation of $1 \cdot 10^{-3}$.

The experiment is repeated 10 times over 10 different data sets, each time the test sequence is 6000 time steps. The results show the mean and standard deviation of the SER over the 10 data sets.

To obtain these results the pump current was set to 187mA. The feedback gain and input gain where optimized for each value of the SNR. The optimal values are in the range -2.35dB to -3.6dB for $\alpha$ and 27.9µW (-15.55dBm) to 55.6µW (-12.55dBm) for $\beta$.

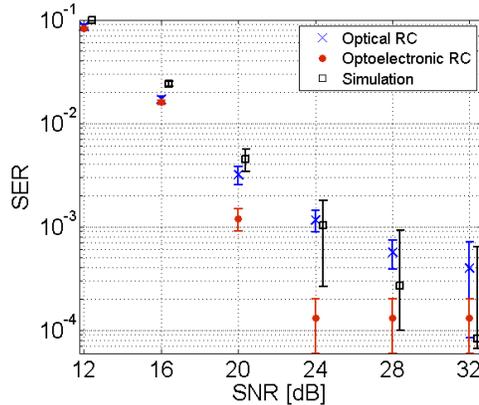

Fig. 5: Performance of the all-optical reservoir on the channel equalization task with 50 nodes at 187mA pump current (Optical RC), with a simulation of the same system (Simulation), and with an optoelectronic reservoir of 50 nodes (OptoElectronic RC). Error bars are statistical.

*4.5. The radar task*

The radar task is a prediction task. Starting from the radar signal backscattered from the ocean surface (collected by the McMaster University IPIX radar [29]), the goal is to predict this signal one to ten time steps in the future. To measure the quality of the prediction, one compares it to the actual data one to ten time steps later. The experiment is conducted for two data sets, termed the low and high sea states. The low sea state corresponds to an average wave height of 0.8 meters (max 1.3 m); the high sea state corresponds to an average wave height of 1.8 meters (max 2.9 m). The inputs, and by consequence the outputs, of the reservoir are two-dimensional, corresponding to the in-phase and in-quadrature outputs (respectively, $I$ and $Q$) of the radar demodulator. For every prediction delay, the first thousand samples of the sea clutter data are used to train our 50 node reservoir, and the next thousand samples are used as test set. The quality of the prediction is measured by the NMSE.

The results presented in Fig. 6, are the best NMSEs for prediction delays ranging from one to ten. Our performances are similar to those of Ref. [19] where the authors have reported, for a reservoir of 80 nodes, a NMSE equal to $1.15 \cdot 10^{-3}$ for one-step predictions and equal to $3.01 \cdot 10^{-2}$ for five-step predictions for low sea state.

The pump current was set to 187mA. The feedback gain $\alpha$ and input gain $\beta$ where optimized for each value of the prediction delay and for each sea state. The optimal values are in the range -2.6dB to -6.6dB for $\alpha$ and 55.6μW (-12.55dBm) to 221.3μW (-6.55dBm) for $\beta$.

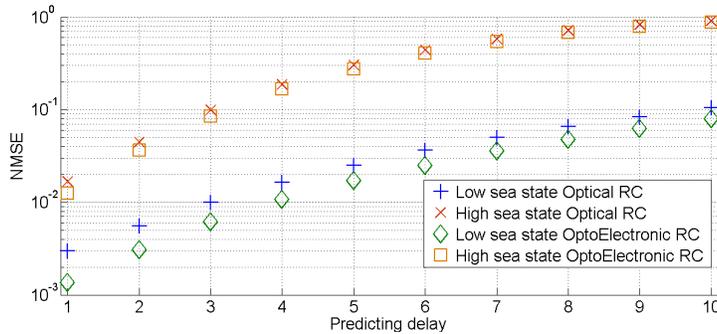

Fig. 6: Predicting radar signal with a reservoir of 50 nodes at a pump current of 187mA (Optical RC) and with the optoelectronic reservoir of 50 nodes reported in Ref. [22] (OptoElectronic RC).

*4.6.Isolated spoken digit recognition*

The isolated spoken digit recognition is a task introduced in Ref. [30] in the context of Reservoir Computing. It consists of the classification of audio sequences, each one containing one digit spoken by one female speaker. The dataset, a subset of the National Institute of Standards and Technology (NIST) TI-46 corpus [31], has 500 sequences: 5 female speakers pronouncing 10 times every digit from 0 to 9.

For each sequence, the audio signal is sampled at 12.5 kHz, preprocessed using the Lyon cochlear ear model [32] and then sampled at regular intervals. The input signal $u(n)$ consists of an 86-dimensional state vector with up to 130 samplings for a single sequence. For this task we used a reservoir of 200 nodes for this task ($N = 200$). The input mask is therefore a *200\*86* matrix $m_{ij}$. Its elements are either -0.1 or +0.1, with equal probability. The masked input driving the reservoir is given by the N-dimensional vector $\sum_j m_{ij} u(n)$.

The output layer consists of 10 linear classifiers $\hat{y}_k(n)$, $k = 0,..,9$, each paired to one digit. For each classifier the desired output is equal to 1 if the corresponding digit is being sent to the reservoir, and to -1 otherwise. For each sequence, the outputs of all the classifiers are averaged over the sequence length; using a winner-take-all approach, the highest averaged classifier is then set to 1, and all other classifiers are set to -1. The metric used to evaluate the recognition is the Word Error Rate (WER), i.e. the rate of digits incorrectly classified.

For this task, since the dataset contains only 500 sequences, the evaluation follows a standard cross-validation process. Five subsets of 100 sequences each are randomly chosen; the reservoir is trained over 4 subsets and tested over the last one. The length of 100 sequences for the test subset ensures that each test subset contains all the digits. This procedure is repeated 5 times rotating the subsets, so that each subset is used once for the test. The results reported here are the average WERs and corresponding standard deviations over the 5 test subsets.

The pump current was set to 270mA. The feedback gain $\alpha$ was set to -7.49 dB and the input gain $\beta$ was set to 55.6μW (-12.55dBm).

The best WER that we can obtain with our all-optical reservoir is 3% (s.d. 1.2%). This is significantly worse than the 0.4% (s.d. 0.55%) WER reported in Ref. [22] for our optoelectronic reservoir. Discrete-time simulations of the same task on our reservoir give perfect classification (0 WER). The difference between the simulations and the experiment is attributed to the ASE noise which is not included in the simulations.

**5.Conclusion**

We presented what is, to the best of our knowledge, the first all-optical implementation of a Reservoir Computer. The system design is based on a simple optical delayed feedback loop combined to the nonlinearity of an optical amplifier, which from a practical viewpoint only requires standard off-the-shelf optical components. We tested our all-optical reservoir on several benchmark tasks previously used in the literature to evaluate reservoir performances. We showed that, despite its simplicity, our system exhibits state-of-the-art capacities for several tasks. This outcome is promising as it reveals the potential of optics in the field of artificial intelligence.

However, compared to the opto-electronic reservoir reported in 22, the performance on tasks, such as the memory capacities, the non linear channel equalization, isolated spoken digit recognition, is somewhat degraded. Our numerical simulations show that this is not due to the form of the nonlinearity which is different in the two experiments, but can be attributed to the noise induced in the present system which arises from the ASE of the SOA. This highlights a difficulty inherent to all analog computation that will have to be addressed in future experimental realizations of reservoir computing.

The configuration studied here is sequential. The reservoir speed is at present limited by the bandwidth of the AWG and digitizer, but it can already process inputs at a rate of 7.8 μs/symbol, regardless of the number of nodes used, even if more time is then needed for the postprocessing. It has moreover the merit of showing that standard optical components with a nonlinear response can be exploited efficiently to perform complex computational tasks. Looking further in the future, it may be possible to exploit this flexibility together with the parallelism of optics in order to perform complex tasks at ultrahigh speed. One possibility, for example, is to encode the state of different nodes at different frequencies, thus implementing reservoirs of higher dimensionality; another is to move to free space optics, and to encode the states of different nodes in the spatial modulation of a light beam. Either way, we hope that the present work will contribute to a renewed development of all-optical computing.

**Acknowledgments**

The authors thank Joni Dambre, Benjamin Schrauwen and Peter Bienstman for helpful discussions. They acknowledge financial support by Interuniversity Attraction Poles program of the Belgian Science Policy Office, under grant IAP P7-35 "photonics@be" and by the Fonds de la Recherche Scientifique FRS-FNRS.